\documentclass[10pt,twocolumn]{article}
\pdfoutput=1

\usepackage[a4paper,top=2cm,bottom=2cm,left=1.2cm,right=1.2cm]{geometry}
\usepackage[english]{babel}
\usepackage[T1]{fontenc}
\usepackage{lmodern}
\usepackage[utf8]{inputenc}
\usepackage[usenames,dvipsnames]{color}
\usepackage{amsmath}
\usepackage{graphicx}
\usepackage{xcolor,calc}
\usepackage[colorlinks]{hyperref}
\hypersetup{linkcolor = black, citecolor = black, urlcolor = black} 
\usepackage{mathptmx}
\usepackage[scaled]{helvet}
\usepackage{fixltx2e}
\usepackage{microtype}
\usepackage{sectsty}
\usepackage[font=small,labelfont={sf,bf}, margin=0cm, indention = 0cm]{caption}
\usepackage{mathbbol}
\usepackage{bm}
\usepackage{soul}
\usepackage{csvsimple}
\usepackage[nomarkers,nolists]{endfloat}
\usepackage{authblk}
\usepackage{rotating}

\usepackage{float}
\usepackage[framemethod=tikz]{mdframed}

\usepackage{etoolbox}

\allsectionsfont{\sffamily\bfseries\normalsize}

\DeclareDelayedFloatFlavor{sidewaystable}{table}
\DeclareDelayedFloatFlavor{sidewaysfigure}{figure}

\usepackage[switch]{lineno}

\begin{document}

\title{\sffamily\bfseries Regional accuracy of ZTE-based attenuation correction in static and 
       dynamic brain PET/MR}

\author[1]{\small Georg~Schramm}
\author[1]{\small Michel~Koole}
\author[1]{\small Stefanie~M.~A.~Willekens}
\author[1]{\small Ahmadreza~Rezaei}
\author[1]{\small Donatienne~Van~Weehaeghe}
\author[2]{\small Gaspar~Delso}    
\author[3]{\small Ronald~Peeters}  
\author[1]{\small Nathalie~Mertens}
\author[1]{\small Johan~Nuyts}    
\author[1]{\small Koen~Van~Laere}

\affil[1]{Department of Imaging and Pathology,
          Division of Nuclear Medicine,
          KU/UZ Leuven, Leuven, Belgium}

\affil[2]{GE Healthcare, Cambridge, UK}

\affil[3]{Division of Radiology
          UZ Leuven, Leuven, Belgium}

\date{}
\maketitle

\newcommand{\petatlas}{PET$_\mathrm{AtlasAC}$}
\newcommand{\petzte}{PET$_\mathrm{ZTEAC}$}
\newcommand{\petct}{PET$_\mathrm{CTAC}$}

\newcommand{\nn}[1]{{\color{black} #1}}
\newcommand{\nnn}[1]{{\color{black} #1}}

\section*{Abstract}
Accurate MR-based attenuation correction (MRAC) is essential
for quantitative PET/MR imaging of the brain.
In this study, we analyze the regional bias caused by MRAC
based on Zero-Echo-Time MR images (ZTEAC) compared to 
CT-based AC (CTAC) in static and dynamic PET imaging.
In addition the results are compared to the performance of
the current default Atlas-based AC (AtlasAC) implemented in the 
GE SIGNA PET/MR.\\
\textbf{\sffamily Methods:}
Thirty static [$^{18}$F]FDG and \nn{eleven} dynamic [$^{18}$F]PE2I acquisitions
from a GE SIGNA PET/MR were reconstructed using ZTEAC (using
a research tool, GE Healthcare), single-subject AtlasAC (the current
default AC in GE's SIGNA PET/MR) and CTAC (from a PET/CT
acquisition of the same day).
In the 30 static [$^{18}$F]FDG reconstructions, the bias
caused by ZTEAC and AtlasAC in the
mean uptake of 85 anatomical volumes of interest (VOIs) of the Hammers' atlas
was analyzed in PMOD.
For the \nn{eleven} dynamic [$^{18}$F]PE2I reconstructions, the bias
caused by ZTEAC and AtlasAC in the non displaceable binding potential BP$_\mathrm{nd}$ 
in the striatum was calculated with cerebellum
as the reference region and a simplified reference tissue model.\\
\textbf{\sffamily Results:}
The regional bias caused by ZTEAC in the static [$^{18}$F]FDG reconstructions 
ranged from -8.0\% to +7.7\% 
(mean 0.1\%, SD 2.0\%). 
For AtlasAC this bias ranged from -31.6\% to +16.6\%
(mean -0.4\%, SD 4.3\%). 
The bias caused by AtlasAC showed a clear gradient in the
cranio-caudal direction (-4.2\% in the cerebellum, +6.6\% 
in the left superior frontal gyrus).
\nn{The bias in the striatal BP$_\mathrm{nd}$ for the [$^{18}$F]PE2I reconstructions ranged from
-0.8\% to +4.8\% (mean 1.5\%, SD 1.4\%) using ZTEAC and from -0.6\% to +9.4\%
using AtlasAC (mean 4.2\%, SD 2.6\%).}\\
\textbf{\sffamily Conclusion:}
ZTEAC provides excellent quantitative accuracy for static and
dynamic brain PET/MR, comparable to CTAC, and is clearly superior to the
default AtlasAC currently implemented in the GE SIGNA PET/MR.

\bigskip

\nn{{\sffamily\textbf{Keywords: MR-based attenuation correction, PET/MR, PET quantification}}}

\section{Introduction}

Since the introduction of combined PET/MR, accurate attenuation correction
(AC) for brain imaging has always been a field of active research.
Neglecting higher bone attenuation of the skull
in the first generation segmentation-based AC methods
used in product implementations led to a substantial spatially-varying 
bias in the reconstructed tracer uptake \cite{Andersen2014}.
To include patient-specific information about higher bone attenuation, two concepts
for MR-based attenuation correction (MRAC) were
investigated by different research groups.
On the one hand, ultra short echo time (UTE) MR sequences that allow to 
generate signal in cortical bone were used to segment bone
structures in the skull 
\cite{Keereman2010,Delso2014,Ladefoged2015,Juttukonda2015,Anazodo2015,Cabello2016}.
On the other hand, the use of single 
\cite{Izquierdo-Garcia2014,Paulus2015,Koesters2016} or multi MR-CT atlas 
\cite{Burgos2014,Merida2015}
information to generate attenuation images including higher bone
attenuation were proposed. 
Recently, Ladefoged et al. \cite{Ladefoged2017} showed in a multi-center
evaluation that the bias introduced by MRAC in
brain PET/MR imaging can be reduced to $\pm$5\% 
when using different second generation
atlas- or UTE-based AC techniques developed by different research 
groups. 

Weiger et al. \cite{Weiger2013} and Wiesinger et al. \cite{Wiesinger2015} 
showed that zero echo time (ZTE) MR sequences have great potential in imaging 
materials with short $T_2^*$ such as cortical bone.
Since ZTE sequences only use a single echo, their
acquisition time is substantially shorter compared to UTE sequences
that usually acquire two echos.
In addition, faster switching from transmit to receive in the ZTE
sequence minimizes loss of signal in tissues with short $T_2^*$ relaxation
times such as cortical bone. 
\nn{Due to the use of minimal gradient switching,
ZTE is less prone to eddy current artifacts than UTE
\cite{Wiesinger2015,Sekine2016b}.}
Moreover, a correlation between the ZTE MR signal intensity and
CT Hounsfield units (HU) in cortical bone was demonstrated in 
\cite{Wiesinger2015}.

Consequently, ZTE MR imaging is very promising for accurate AC
in brain PET/MR.
Delso et al. \cite{Delso2015} showed that
ZTE-based skull segmentation, which is needed to generate
attenuation images including higher bone attenuation, is feasible.
Boydev et al. \cite{Boydev2017} showed that the use of
ZTE MR images in their atlas-based prediction of pseudo CTs improved
the correctness of the pseudo CTs for radiation therapy
planning in case of bone resection surgery prior to the 
radiation therapy compared to using T1-weighted MR images
as input. 

Moreover, Sekine et al. \cite{Sekine2016b}, Khalife et al. 
\cite{Khalife2017}, Yang et al. \cite{Yang2017}, Leynes et al.
\cite{Leynes2017}, \nn{Wiesinger at al \cite{Wiesinger2018}} 
recently demonstrated that the quantitative 
accuracy of PET images reconstructed with ZTE-based attenuation 
images is high.
All groups investigated pilot studies with small patient cohorts
(10, 16, 12, 6, and 5 subjects, respectively) and evaluated static FDG PET images.

So far, no evaluation of ZTE-AC for absolute quantification 
of dynamic receptor studies (e.g. in terms of non-displaceable binding potential
or distribution volume) has been published.
The influence of attenuation correction on parameters derived
from kinetic modeling is more complex especially in case when
reference tissue models are used.
In those cases, it is important to have accurate attenuation
correction for the target region (e.g. the striatum) as well
as for the reference region (e.g. the cerebellum).

To study the influence of ZTE-based AC
on the accuracy of tracer kinetic modeling using the
simplified reference tissue model, we analyzed
\nn{eleven} dynamic PET/MR acquisitions with the
highly selective dopamine transporter tracer [$^{18}$F]PE2I.
In addition, we investigated the regional 
quantitative accuracy of ZTE-based AC
in 30 static [$^{18}$F]FDG PET/MR acquisitions. 

\section{Materials and Methods}

\subsection{Ethical approval and informed consent}
All procedures performed in studies involving human participants were 
in accordance with the ethical standards of the institutional 
and/or national research committee and with the 1964 Helsinki 
declaration and its later amendments or comparable ethical standards.
Informed consent was obtained from all individual 
participants included in the study.

\subsection{Subjects}
We included \nn{48} subjects that participated in two ongoing PET/MR
research protocols in the context of neurodegenerative 
diseases. 
Thirty-four patients suspected for dementia were investigated 
with a static [$^{18}$F]FDG PET/MR
protocol between October 2016 and June 2017.
Three  patients were excluded from this comparison study due to dental implants
which led to metal artifacts in the MR images.
In one case the patient was positioned too low in
the head coil which led to very low ZTE MR signal in
the caudal end of the head due to low coil sensitivity
in that region.
This case was excluded as well.
The mean age of the remaining 30 patients was 63\,y
(range 40-77\,y).
In addition, we analyzed \nn{14} dynamic [$^{18}$F]PE2I acquisitions
of healthy controls (mean age 40.8\,y, range 21-70\,y).
As in the case of the static acquisitions, three cases had to be
excluded due to metal artifacts caused by dental implants. 

\subsection{Imaging protocol}
All patients were examined on a GE SIGNA TOF PET/MR (GE Healthcare, Chicago, US) 
\cite{Grant2016}.
The static [$^{18}$F]FDG PET/MR protocol included a 25\,min
static PET acquisition $66 \pm 9$\,min after tracer injection
(mean injected dose $144 \pm 31$\,MBq).
For the \nn{eleven} [$^{18}$F]PE2I cases, 60\,min of dynamic PET data
were acquired directly after tracer injection (mean injected dose $153 \pm 15$\,MBq). 
During the PET acquisitions a LAVA flex MR 
(acquisition details: repetition time 4\,ms, echo time 2.23\,ms, flip angle\,5$^\circ$, 
matrix 256\,x\,256\,x\,120, voxel size 1.95\,mm x 1.95\,mm x 2.6\,mm, number of averages 0.7, 
acquisition time: 18\,s),
a ZTE MR (acquisition details:
3D radial acquisition, flip angle 0.8$^\circ$, matrix 110\,x\,110\,x\,116, 
voxel size 2.4\,mm x 2.4\,mm x 2.4\,mm, number of averages 4, 
bandwidth $\pm$ 62.5\,kHz, acquisition time 42\,s) 
and other study-specific MR sequences were acquired.
Among the study-specific MR sequences were a 
3D volumetric sagittal T1-weighted BRAVO sequence 
(acquisition details: echo time 3.2\,ms, repetition time 8.5\,ms, 
inversion time 450\,ms, flip angle 12$^\circ$, receiver bandwidth $\pm$ 31.2\,kHz,
NEX 1, voxel size 1\,mm x 1\,mm x 1\,mm) 
and a 3D sagittal T2-weighted CUBE FLAIR sequence 
(acquisition details:
echo time 137\,ms, number of echoes 1, repetition time 8500\,ms, 
inversion time 50\,ms, receiver bandwidth $\pm$31.25\,kHz, NEX 1, 
voxel size 1.2\,mm x 1.3\,mm x 1.4\,mm)
\nnn{In all cases, a standard head coil (8-channel HR brain, GE Healthcare) 
was used for the MR acquisitions.}

All subjects underwent a PET/CT acquisition
before ([$^{18}$F]FDG cases) or after ([$^{18}$F]PE2I cases) 
the PET/MR acquisition.
The PET/CT acquisitions were performed on a Siemens Biograph 16
or with a Siemens Biograph 40 (Siemens Healthcare, Erlangen, Germany) PET/CT. 
All PET/CT examinations included a low-dose CT acquisition
(120\,kV, 11\,mAs) which was used to generate 
a CT-based attenuation image taken as the ground truth in the study.
  
\subsection{PET image reconstruction}
\begin{figure*}
  \centering
  \includegraphics[width = 0.7\textwidth]{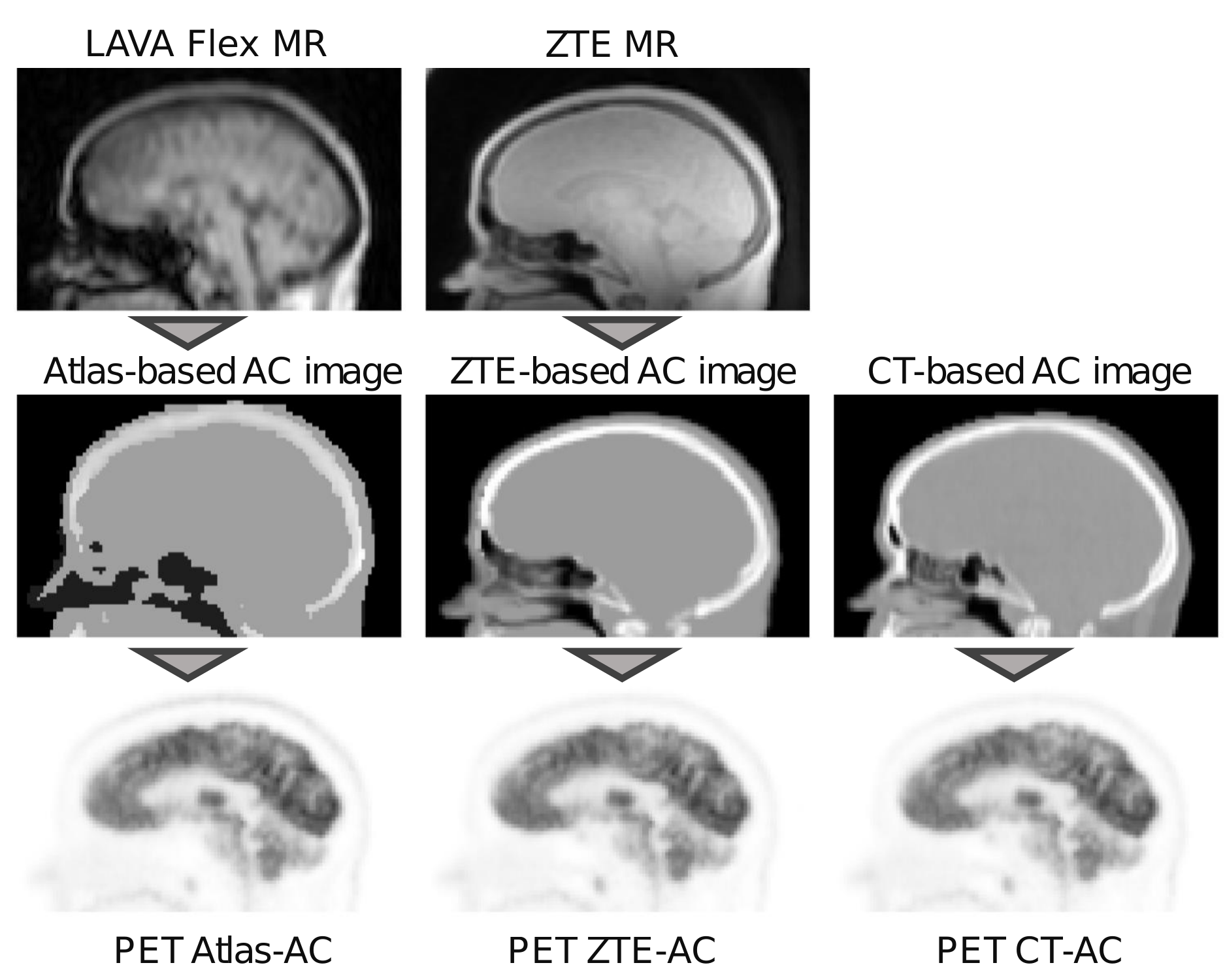}
  \caption{Workflow of the PET reconstructions used in this study.
           In the reconstruction of {\petatlas} an atlas-based attenuation
           image that was derived from a LAVA flex MR image was used
           for attenuation correction (left column).
           The atlas-based attenuation image was generated with the
           vendor-provided software that is used in clinical routine.
           In the reconstruction of {\petzte}
           attenuation correction was performed using
           a ZTE-based attenuation image that was derived from
           a ZTE MR image (middle column).
           The ZTE MR post-processing was done with a research tool
           provided by the vendor.
           For {\petct} a coregistered CT-based attenuation image
           from a PET/CT acquisition of the same day was used.
           In all attenuation images, templates for the bed and
           the head coil were added.}
  \label{fig:workflow}
\end{figure*}

The PET raw data from all PET/MR acquisitions were reconstructed
with three different methods for attenuation correction,
\nn{shown in Figure~\ref{fig:workflow}.}
First, a GE atlas-based attenuation image (current default method
in the SIGNA PET/MR) was used to reconstruct {\petatlas}.
Subsequently, a GE ZTE-based attenuation image and a coregistered 
CT-based attenuation image were used to reconstruct
{\petzte} and {\petct}, respectively. 
The generation of all attenuation images is described in detail
in the following subsection.
The reconstructions of the static PET data sets were performed offline
with the GE reconstruction toolbox v.1.28 (GE Healthcare, Chicago, US) 
using time of flight ordered subset maximum likelihood expectation maximization 
(TOF OSEM) with 4 iterations and 28 subsets, a voxel size of 
1.17\,mm x 1.17\,mm x 2.78\,mm, and a Gaussian post-smoothing with an FWHM of 4\,mm.

Reconstruction of the dynamic [$^{18}$F]PE2I PET data sets was performed
on the scanner (software version MP24.R03).
The acquired listmode data were split into 32 frames (frame length
10\,s to 360\,s).
All frames were reconstructed with TOF OSEM with 4 iterations 
and 28 subsets, a voxel size of 1.56\,mm x 1.56\,mm x 2.78\,mm 
and a Gaussian post-smoothing with an FWHM of 3\,mm. 

\subsection{Generation of attenuation images}
First, the atlas-based attenuation images were generated 
with the GE reconstruction toolbox v.1.28 which uses
the same post-processing algorithm as 
\nn{implemented in the current software release 
of the SIGNA PET/MR (MP24.R03)}.
The algorithm uses a non-rigid registration of an
input in-phase LAVA flex MR image to an atlas of predefined
attenuation images \cite{Sekine2016,Sekine2016a}.
\nnn{The resulting atlas-based attenuation images are post-smoothed
with a Gaussian kernel with FWHM ca. 10\,mm.} 

Second, the ZTE-based attenuation images were generated by 
post-processing the ZTE MR images with a research tool
provided by GE (v.1.6.2).
The upcoming software release of the SIGNA PET/MR (MP 26) 
will contain an option \nn{to use this algorithm} for ZTE-based AC.
The ZTE post-processing algorithm identifies bone voxels
based on the ZTE image intensity and assigns continuous
bone attenuation values \cite{Wiesinger2015,Delso2015,Sekine2016b}.
The bone segmentation in the ZTE post-processing is completely
model-free.
\nn{To avoid missclassifications of air, tissue and bone in the
nasal region, the ZTE post-processing algorithm v.1.6.2 uses
the sinus/edge correction evaluated in \cite{Yang2017}}.

The ZTE post-processing provided by GE has several input
parameters. 
For all parameters but one (the partial volume slope) we used
the default values suggested by GE.
We used a value of 2 for the parameter for the partial volume 
slope which was obtained based on an evaluation of the
results of the first 15 static subjects.
\nnn{The main influence of the partial volume slope parameter that 
we observed was a change in the size of the outer contour of the head 
(transition between background air and soft tissue of the skin).
By changing the partial volume slope we obtained better agreement 
with the size of the outer contours derived from the CT-based 
attenuation images.
When using the default partial volume slope of 1, the outer contour of the head
is dilated by 1 voxel (2.4\,mm) compared to using a partial volume slope of 2.
This in turn led to a small global positive bias of 3\%.}
\nnn{All ZTE-based attenuation images were post-smoothed with a
Gaussian kernel with FWHM 4\,mm.}

Third, after automatically removing the patient bed and cushions
the low-dose CT images from the PET/CT acquisition
were rigidly coregistered to the in-phase LAVA flex MR.
Subsequently, the Hounsfield units of the coregistered CT
were scaled to 511-keV attenuation coefficients by using the
GE-provided multi-linear scaling. 
\nnn{We verified that the Siemens and GE scaling for 120 kV are virtually 
identical up to 1200 HU (where GE decreases the slope while Siemens does not).}
After adding the templates for the PET/MR patient table and
the head coil, a CT-based attenuation image that could be
used to reconstruct the PET/MR raw data was obtained.
\nnn{All CT-based attenuation images were also post-smoothed with a
Gaussian kernel with FWHM 4\,mm.}

The axial field of view (FOV) of the ZTE MR
(limited by the sensitivity of the head coil)
and the one of the CT was slightly smaller
than the axial FOV of the PET detector rings in the SIGNA PET/MR.
To complete areas in the neck and shoulders where ZTE or 
CT image information was not available,
a simple segmentation-based two class 
attenuation image based on the LAVA flex MR image was used.

\subsection{Image analysis of static acquisitions}

For all static acquisitions the mean uptake in 85 anatomical
volumes of interest (VOIs) was calculated in {\petatlas}, 
{\petzte}, and {\petct}.
The VOIs were defined in the neuro tool of PMOD v.3.8 
(PMOD technologies LCC, Zurich, Switzerland)
using the Hammers atlas \cite{Hammers2003}.
In every VOI we calculated the fractional bias of the mean
uptake as

\begin{align}
b_\mathrm{AtlasAC}(\mathrm{VOI}) &= 
\frac{a_\mathrm{AtlasAC}(\mathrm{VOI}) - a_\mathrm{CTAC}(\mathrm{VOI})}{a_\mathrm{CTAC}(\mathrm{VOI})} \label{eq:batlas} \\ 
b_\mathrm{ZTEAC}(\mathrm{VOI})   &= 
\frac{a_\mathrm{ZTEAC}(\mathrm{VOI}) - a_\mathrm{CTAC}(\mathrm{VOI})}{a_\mathrm{CTAC}(\mathrm{VOI})} \label{eq:bzte} ,
\end{align}
where $a_\mathrm{CTAC}(\mathrm{VOI})$ is the mean uptake of the VOI in {\petct} that was used
as the gold standard and $a_\mathrm{AtlasAC}(\mathrm{VOI})$ and $a_\mathrm{ZTEAC}(\mathrm{VOI})$ 
are the mean uptake of the VOI in {\petatlas} and {\petzte}, respectively.
In three subjects (6,7, and 21), the caudal end of the
occipital skull was not completely in the FOV in the attenuation
CT.
In those subjects, the cerebellum VOIs were excluded from
the analysis.
All VOIs were grouped according to their anatomical location into the following
groups: frontal cortex, temporal cortex, parietal cortex, occipital cortex,
medial temporal cortex, striatum, thalamus, cerebellum, and cerebral white matter.
All VOIs and the assigned groups (regions) are listed in \nn{supplementary} Tables~\ref{tab:regional_bias}
and \ref{tab:regional_biastwo}.
A Wilcoxon signed-rank test was used to test whether the subject avaraged mean
of $a_\mathrm{AtlasAC}$ and $a_\mathrm{ZTEAC}$ is different from $a_\mathrm{CTAC}$
in all VOIs and regions.

To analyze the robustness of GE's atlas-based and ZTE-based attenuation correction, 
we applied the metric proposed in the multi-center evaluation of Ladefoged et al.
\cite{Ladefoged2017}. 
This metric calculates the fraction of subjects in which the MRAC-introduced voxel bias
of at least a given fraction of brain voxels is within $\pm 5\%, \pm 10\%, \pm 15\%$.
As mentioned in \cite{Ladefoged2017}, for a perfect AC method 100\% of the subjects
100\% of the brain voxels would be within $\pm 0\%$.
The results of this metric were visualized in a characteristic curve for the three 
bias thresholds $\pm 5\%, \pm 10\%, \pm 15\%$.
As in \cite{Ladefoged2017}, we also analyzed three subjects with the biggest fraction
of voxels exceeding a bias of $\pm 10\%$.

\subsection{Image analysis of dynamic acquisitions}

Regional time activity curves (TACs) were extracted for the
left and right caudate nucleus, left and right putamen, 
and the cortex of the cerebellum.
All VOIs were defined based on the 3D T1 BRAVO MR image using the Freesurfer 
image analysis suite which is documented
and freely available online (http://surfer.nmr.mgh.harvard.edu/)
\cite{Fischl2002}.
Subsequently, we used the simplified reference tissue model (SRTM)
with the cerebellar grey matter as reference region to estimate binding potential values 
(BP$_\mathrm{nd}$) in the four striatal VOIs.

\nn{As proposed in \cite{Lammertsma1996} and validated for [$^{18}$F]PE2I
in \cite{Sasaki2012}, the tissue response $C_t(t)$ was modeled as
\begin{equation}
C_t(t) = R_1 C_r(t) + \left(k_2 - \frac{R_1\,k_2}{1 + \mathrm{BP_{nd}}} \right) C_r(t)
         * \exp \left( \frac{-k_2\,t}{1 + \mathrm{BP_{nd}}} \right) \ ,
\label{eq:srtm}
\end{equation}
where $C_r(t)$ is the TAC of the reference tissue (the cerebellum), $R_1$ is the ratio
between $K_1$ of the tissue and reference tissue, and $*$ denotes the convolution
operator.
The parameters $R_1$, $k_2$, and BP$_\mathrm{nd}$ were obtained with non-linear curve fitting
using the python package lmfit (v.0.9.7).}

In a similar way to Eqs.~(\ref{eq:batlas}) and (\ref{eq:bzte}), we calculated
the bias of BP$_\mathrm{nd}$, \nn{$R_1$, and $k_2$} in the four striatal VOIs for 
{\petatlas} and {\petzte} compared to {\petct}.
In all VOIs, a Wilcoxon signed-rank test was used to test whether the subject averaged mean of
BP$_\mathrm{nd,AtlasAC}$ and BP$_\mathrm{nd,ZTEAC}$ differ from BP$_\mathrm{nd,CTAC}$.

\section{Results}

\subsection{Regional bias in static PET imaging}

\begin{figure*}
  \centering
  \includegraphics[width = 0.9\textwidth]{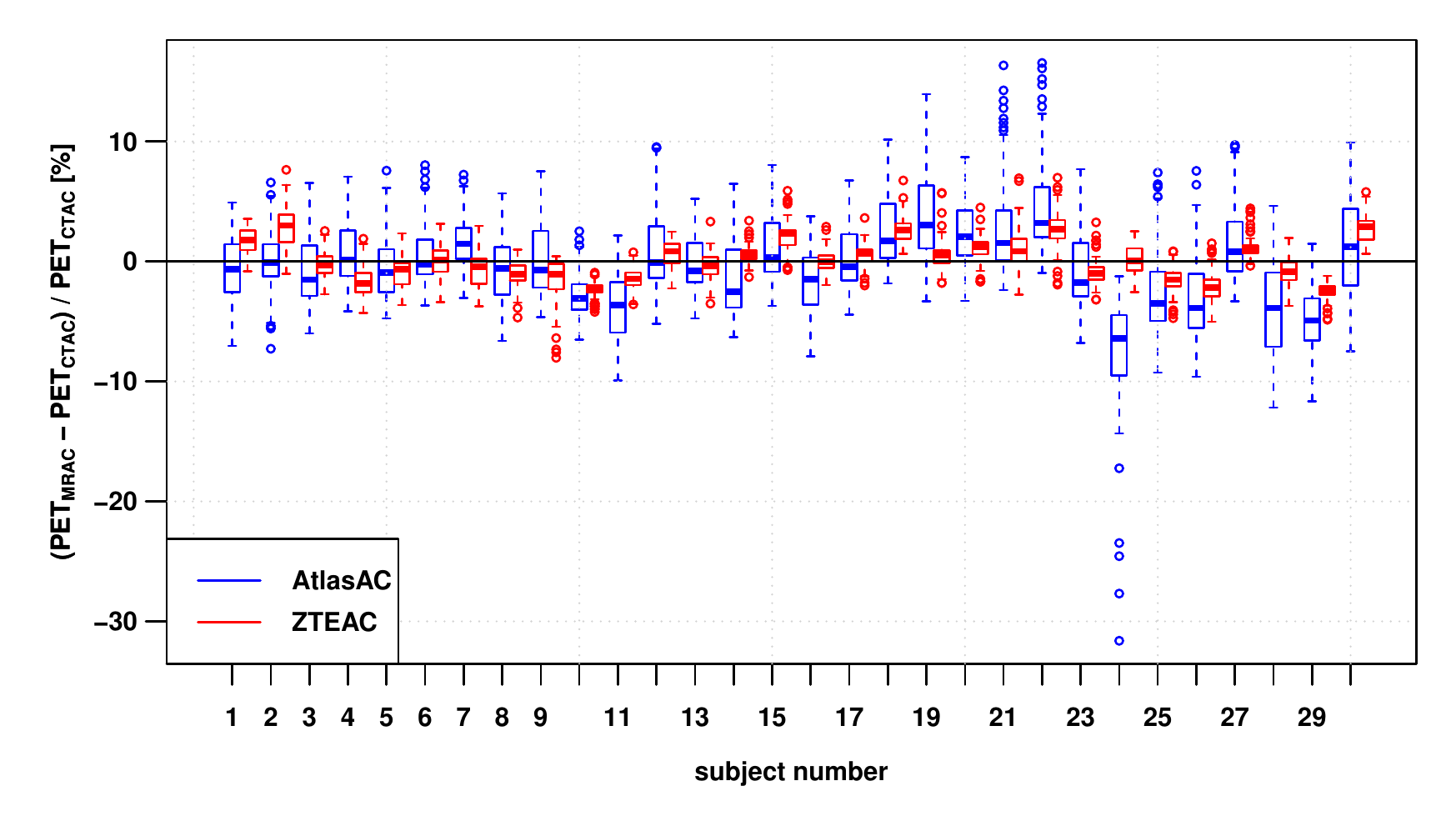}
  \caption{Regional bias in the PET reconstruction caused by  
           AtlasAC (blue) and ZTEAC (red) compared
           to CTAC for all 30 static PET acquisitions.
           Each box plot shows the bias distribution over the 85
           anatomical VOIs.
           The rectangular boxes represent the interquartile ranges (IQR)
           and the horizontal line are the medians. The upper ends of the
           whiskers are at the minimum of: the third quartile plus 1.5IQR
           and the biggest data point 
           The lower ends of the
           whiskers are at the maximum of: the first quartile minus 1.5IQR
           and the smallest data point.
           Outliers are plotted with open circles.
           \nn{Please note that the VOIs in the cerebellum had to be excluded in 
           3 subjects (6,7,21).}}
  \label{fig:subjects_box}
\end{figure*}

\begin{figure*}
  \centering
  \includegraphics[width = 0.9\textwidth]{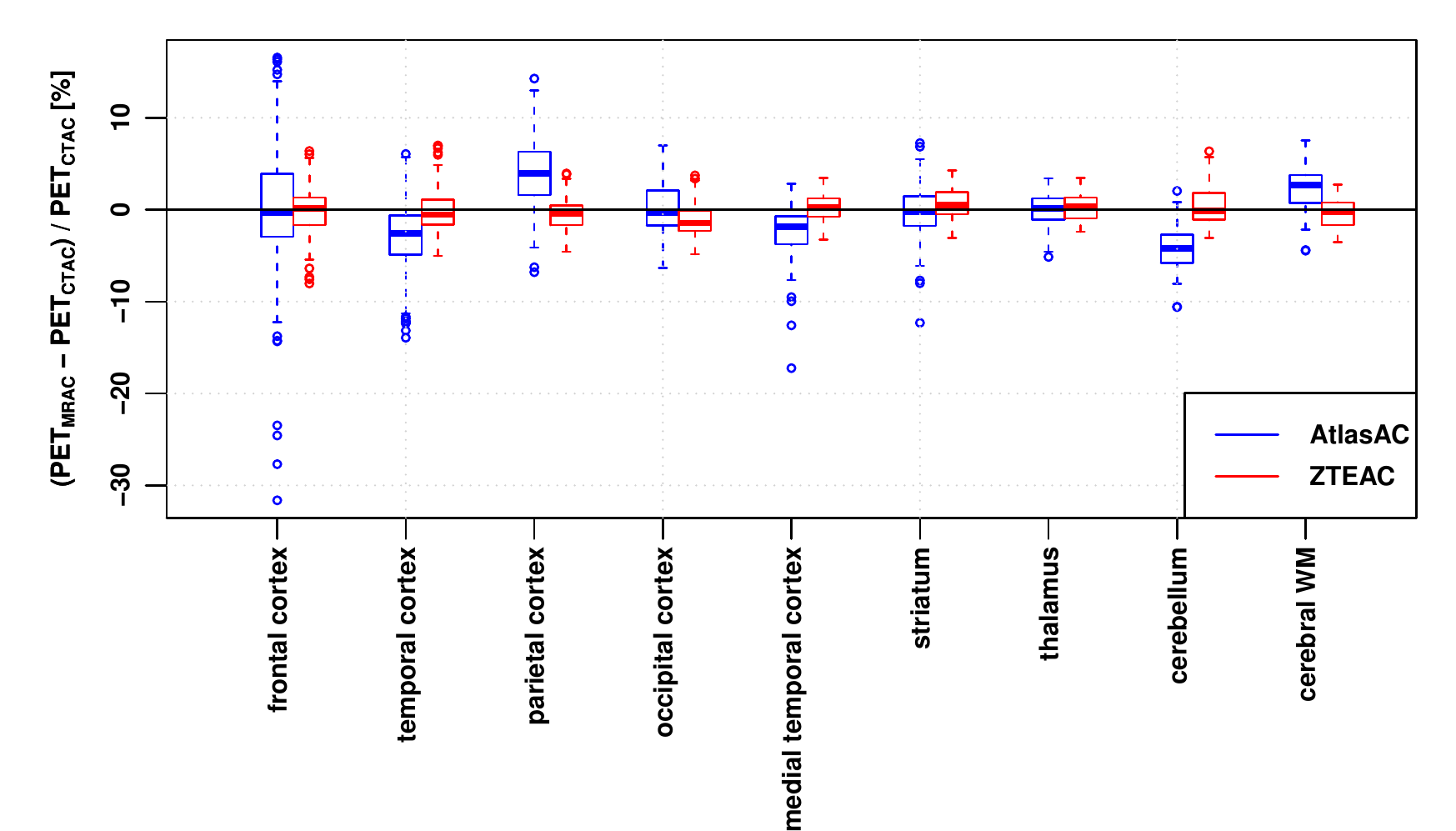}
  \caption{Regional bias in the PET reconstruction caused by  
           AtlasAC (blue) and ZTEAC (red) compared
           to CTAC as a function of the VOI location in the brain.
           \nn{Please note that the VOIs in the cerebellum had to be excluded in 
           3 subjects (6,7,21).}}
  \label{fig:regions_box}
\end{figure*}

\begin{figure*}
  \centering
  \includegraphics[width = 0.9\textwidth]{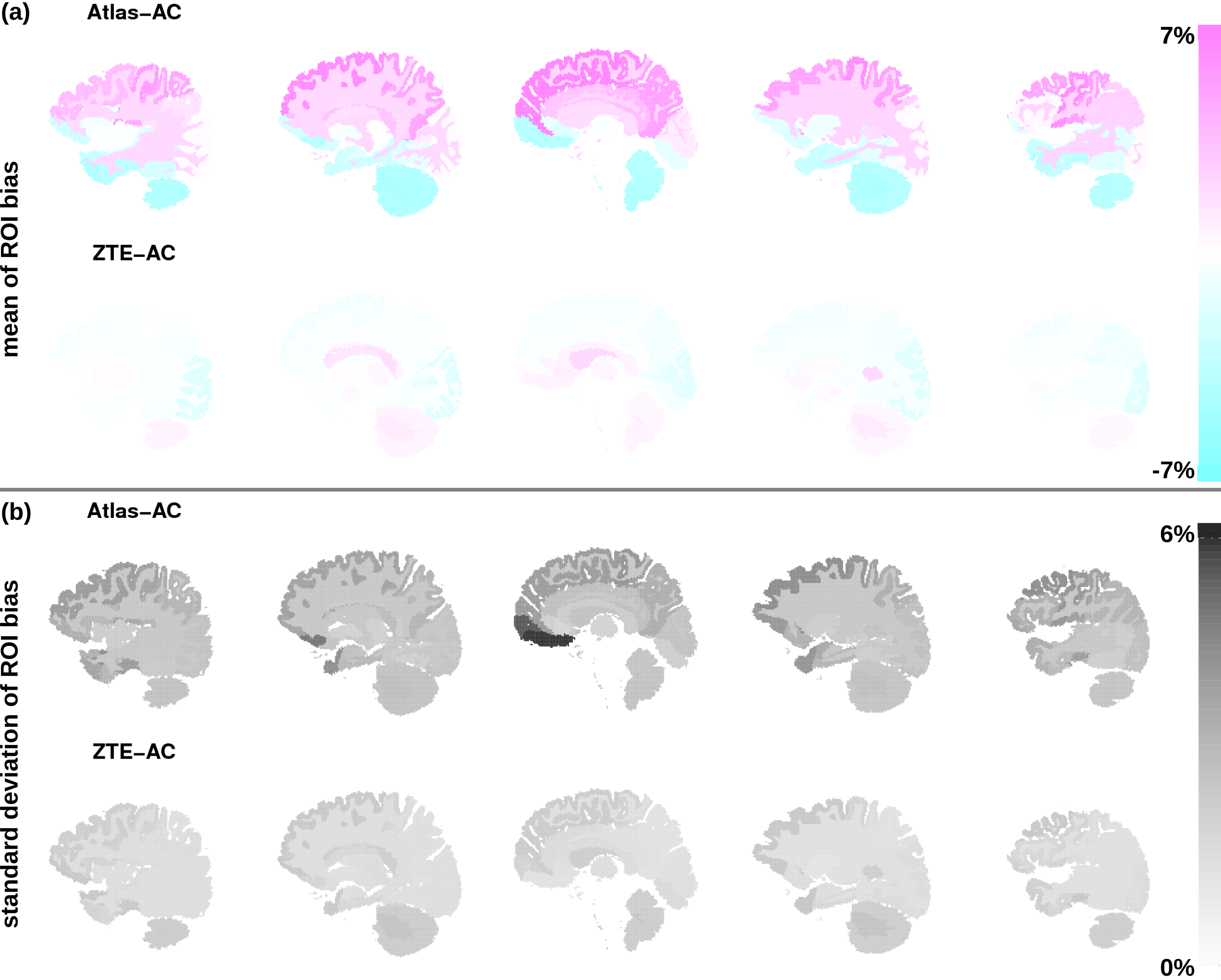}
  \caption{(a) mean of bias in the PET reconstruction in all 85 anatomical
           VOIs averaged over all 30 static PET acquisitions for Atlas-AC
           (top row) and ZTE-AC (bottom row).
           (b) standard deviation of VOI bias.
           The VOI location is visualized in five \nn{sagittal} slices using the brain anatomy
           of subject 1.
           \nn{Please note that the VOIs in the cerebellum had to be excluded in 
           3 subjects (6,7,21).}}
  \label{fig:regional_bias}
\end{figure*}

\begin{figure*}
  \centering
  \includegraphics[width = 0.8\textwidth]{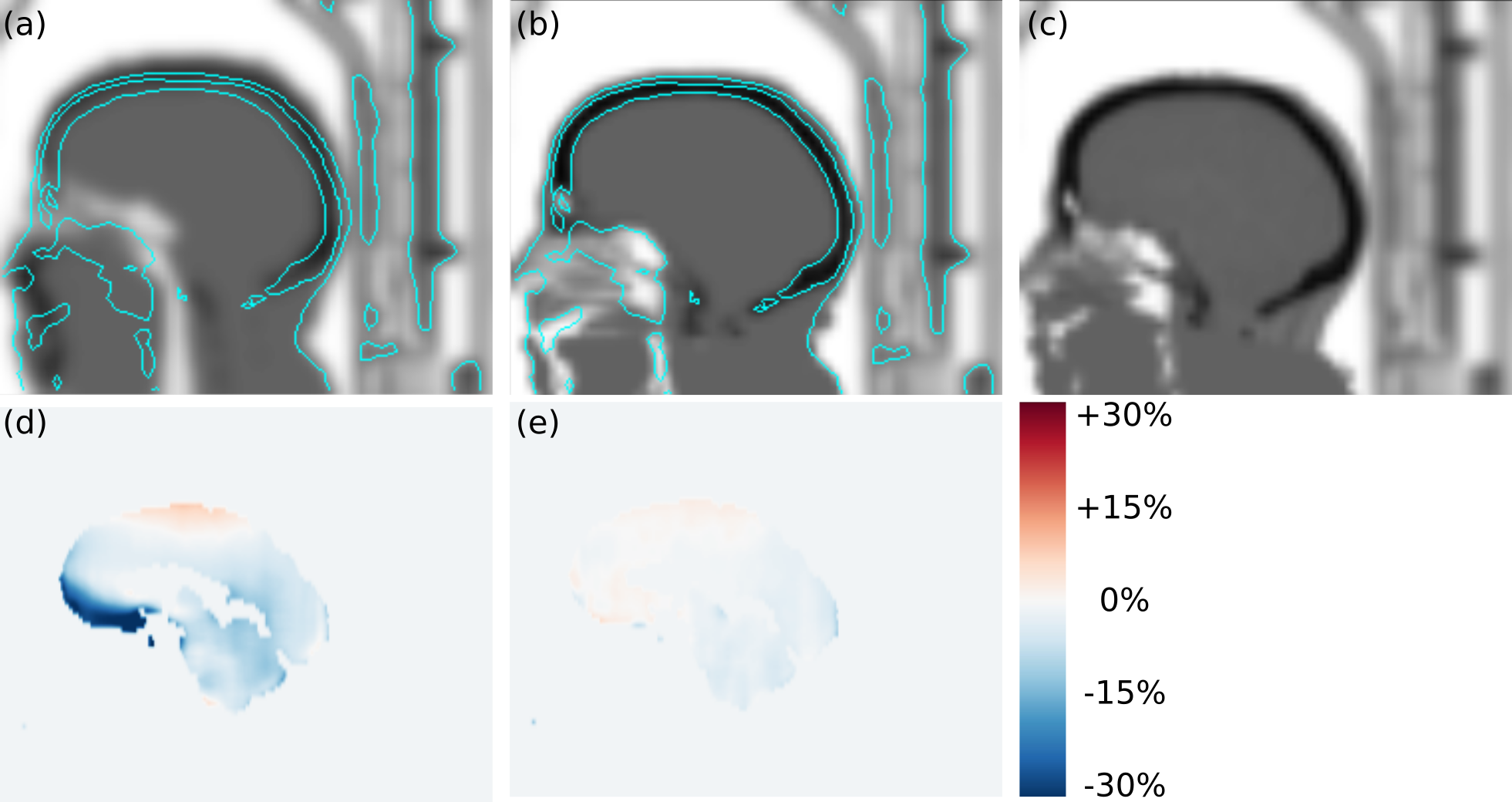}
  \caption{Transversal slices of (a) atlas-based attenuation image, (b) ZTE-based 
           attenuation image, (c) CT-based attenuation image, (d) regional bias 
           in {\petatlas}, and (e) regional bias in {\petzte}
           of subject 24.
           In this case, the template registration in the atlas-based attenuation 
           image failed which caused a misclassification of soft tissue voxels as 
           air voxels in the frontal region.
           The cyan contour lines show the head contour in the CT-based attenuation 
           image for comparison.
           As a result of the underestimated attenuation {\petatlas} shows strong 
           negative bias of up to -32\% in the left straight gyrus.}
  \label{fig:patex}
\end{figure*}

\nn{Figures~\ref{fig:subjects_box} and \ref{fig:regions_box}, 
and supplementary Tables~\ref{tab:subjects}, and \ref{tab:region} show the 
results for the regional bias in the 
static [$^{18}$F]FDG reconstructions caused by ZTEAC and AtlasAC compared to CTAC
on a subject and regional level, respectively.
Globally the bias ranges from ranges from -31.6\% to +16.6\% with
a mean of -0.4\% and a standard deviation of 4.3\% for {\petatlas}.
For {\petzte} the bias ranges from -8.0\% to +7.7\% with
a mean of 0.1\% and a standard deviation of 2.0\%.
Excluding the outliers based on the boxplot shown in Fig.~\ref{fig:regions_box}
reduces the global bias range to -12\% to +14\% for {\petatlas} and
to -5.5\% to +5.5\% for {\petzte}.

On a subject level, Fig.~\ref{fig:subjects_box} and supplementary Table~\ref{tab:subjects} 
demonstrate that ZTEAC strongly reduces the inter- and intra-subject variability in the bias.
In subject 24 where the non-rigid alignment to the atlas failed,
{\petatlas} showed severe negative bias of more than -25\% in the 
orbitofrontal cortical VOIs (see Fig.~\ref{fig:patex}).
In these VOIs of subject 24, the bias of {\petzte}
was less than 1.6\%.

On a regional level, Fig.~\ref{fig:regions_box} and supplementary Table~\ref{tab:region} 
show that ZTEAC strongly reduces the inter- and 
intra-regional variability in the bias, as well as the mean bias \nnn{in the frontal cortex, 
temporal cortex, parietal cortex, medial temporal cortex, cerebellum, and cerebral white matter.}
In all regions shown in Fig.~\ref{fig:subjects_box}, the mean bias in {\petzte} 
is between -1.2\% and +0.6\%.
{\petatlas} shows a distinct negative bias in the cerebellum (mean -4.2\%) and distinct
positive bias in the parietal cortex (mean +4\%).

On a VOI level, Fig.~\ref{fig:regional_bias} and supplementary Tables~\ref{tab:regional_bias} 
and Tables~\ref{tab:regional_biastwo} demonstrate that ZTEAC strongly reduces the inter- and 
intra-VOI variability in the bias,
as well as the mean bias in almost all VOIs.
The mean VOI bias caused by ZTEAC ranges from -1.8\% in the 
lateral remainder of the left occipital lobe to +2.2\% in the left 
lateral ventricle.
In {\petatlas}, a distinct gradient in the mean VOI bias in the
cranio-caudal direction is visible.
The mean VOI bias caused by AtlasAC ranges from -4.5\% in the 
cerebral white matter to +6.6\% in the left superior frontal gyrus.  
In {\petzte}, only 1.4\% of the analyzed VOIs in all subjects 
had a bias of more than 5\% whereas in {\petatlas} 20.3\% of 
all VOIs showed a bias of more than 5\%.}

\begin{figure*}
  \centering
  \includegraphics[width = 0.5\textwidth]{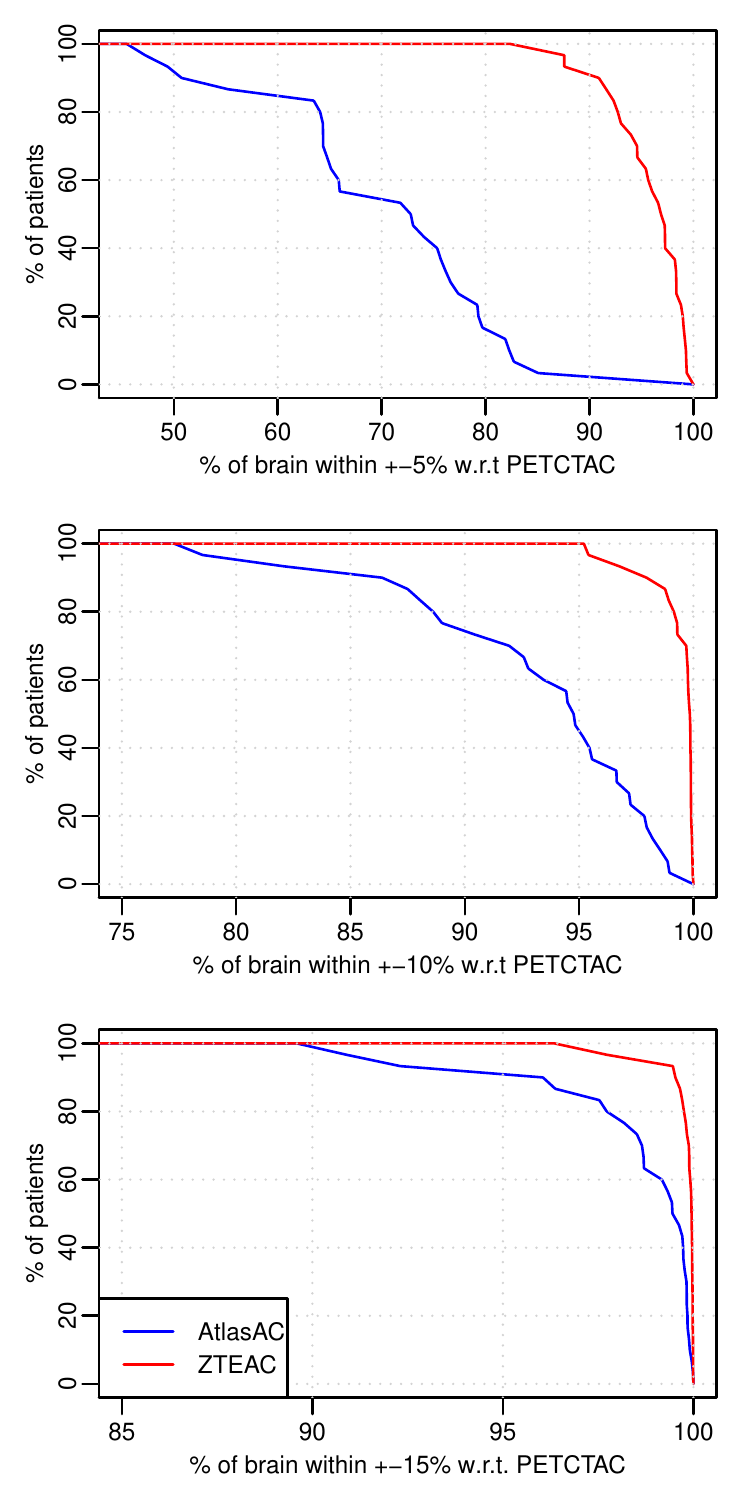}
  \caption{Outlier analysis \cite{Ladefoged2017} of the 30 static acquisitions
           for {\petzte} (red) and {\petatlas} (blue). 
           Note the different scale on the x-axis.}
  \label{fig:outlier}
\end{figure*}

Figure~\ref{fig:outlier} shows the results of the outlier metric
\cite{Ladefoged2017} for biases within
($\pm 5\%, \pm 10\%, \pm 15\%$).
Again, the performance of ZTE-based attenuation correction
is much better than the one of the atlas-based attenuation
correction.
At least 95\% / 77\% of all brain voxels in all subjects show 
a bias within $\pm 10\%$ for {\petzte} / {\petatlas}.  
For a bias within $\pm 5\%$ the corresponding values are
82\% / 46\% and 
for a bias within $\pm 15\%$ the corresponding values are
97\% / 89\%.
Table~\ref{tab:outlier} shows results of three worst outliers
in terms of subjects with highest voxel bias, highest VOI bias and 
highest fraction of brain exceeding a bias of $\pm$10\%.

\begin{table*}
\centering
\footnotesize
\sffamily
\caption{Results of outlier analysis of the static acquisitions in terms of subjects with highest 
         voxel bias, highest VOI bias and 
         highest fraction of brain exceeding a bias of $\pm$10\%}
\label{tab:outlier}
\begin{tabular}{llll}
 & \multicolumn{3}{c}{\textbf{subjects with highest single voxel bias}}                                       \\
AtlasAC & subject 24                  & subject 29                          & subject 21                       \\
        & -53\% (left straight gyrus) & -43\% (left anterior temporal lobe) & +41\% (left superior frontal gyrus)\\
ZTEAC   & subject 5                   & subject 18                    & subject 21                            \\
        & +45\% (left cerebellum)     & +43\% (left cerebellum)       & +41\% (right fusiform gyrus)          \\
\noalign{\vskip 5mm}
 & \multicolumn{3}{c}{\textbf{subjects with highest VOI bias}}                                                \\
AtlasAC & subject 24                  & subject 22                            & subject 21                    \\
        & -31\% (left straight gyrus) & +16\% (left suprerior frontal gyrus)  & +16\% (left precentral gyrus) \\
ZTEAC   & subject 9                   & subject 22                            & subject 21                    \\
        & -8\% (left middle frontal gyrus) & +7\% (right superior temporal gyrus) & +7\% (right fusiform gyrus)\\
\noalign{\vskip 5mm}
 & \multicolumn{3}{c}{\textbf{subjects with highest fraction of brain exceeding a bias of $\pm$10\%}}         \\
AtlasAC & subject 22                  & subject 19                         & subject 21                       \\
        & 23\%                        & 22\%                               & 18\%                             \\
ZTEAC   & subject 21                  & subject 6                          & subject 9                        \\
        & 5\%                         & 5\%                                & 3\%
\end{tabular}
\end{table*}

\subsection{Bias in kinetic modeling of [$^{18}$F]PE2I}

\begin{figure*}
  \centering
  \includegraphics[width = 0.95\textwidth]{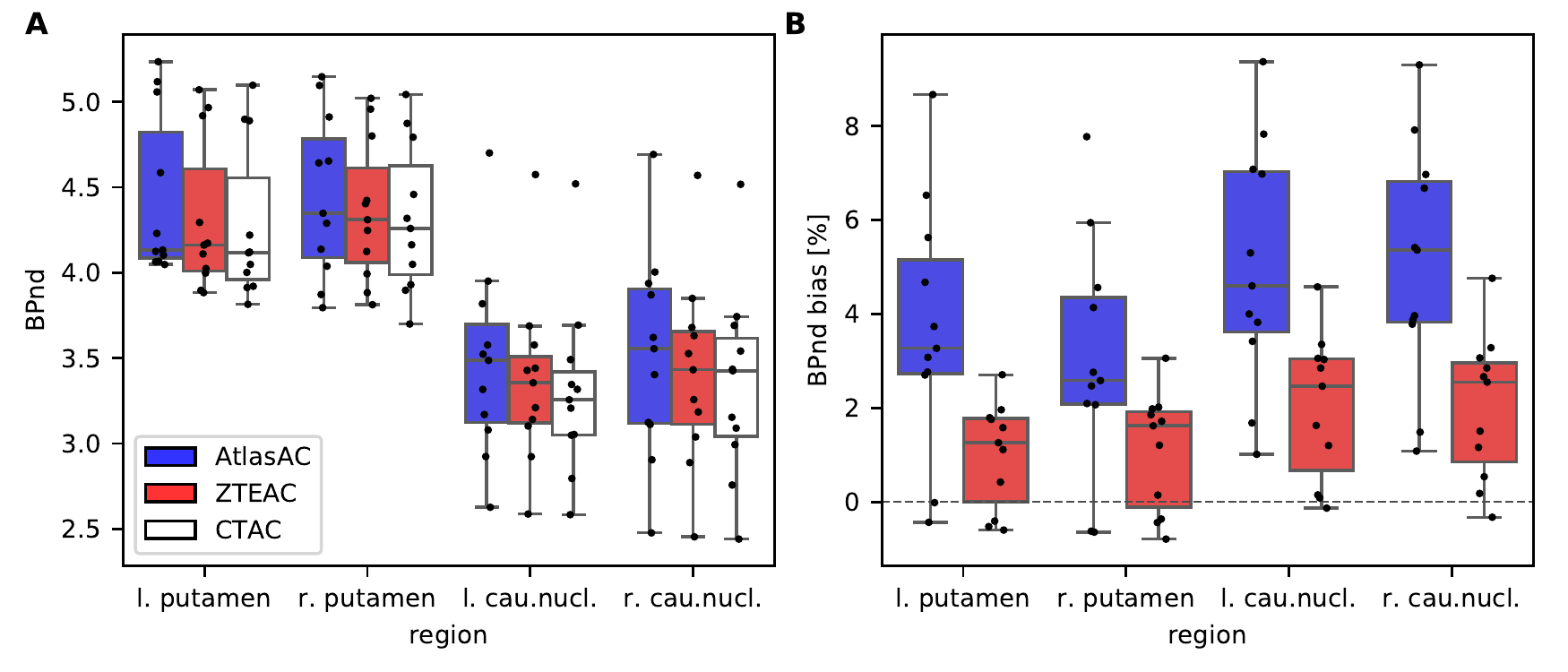}
  \caption{\nn{
           (A) Boxplot of BP$_\mathrm{nd}$ values in four striatal regions of 
               the 11 [$^{18}$F]PE2I subjects 
               obtained from {\petatlas}, {\petzte}, and {\petct}.   
           (B) Bias in BP$_\mathrm{nd}$ estimation in {\petatlas}, {\petzte} compared to {\petct}.}}
  \label{fig:BPnd_bias}
\end{figure*}

\nn{Figure \ref{fig:BPnd_bias}} and supplementary Table~\ref{tab:bpbias} summarize the bias in 
the modeled BP$_\mathrm{nd}$  in four different regions of the striatum using the cerebellum as 
reference region and TACs derived from {\petatlas} and {\petzte}
compared to TACs from {\petct}.
\nn{The bias in the BP$_\mathrm{nd}$ ranges from -0.6\% (right putamen in subject 11) 
to +9.4\% (left caudate nucleus subject 9) for {\petatlas} and 
from -0.8\% (right putamen subject 3) to +4.8\% (right caudate nucleus subject 8) for {\petzte}.
The right caudate nucleus shows the biggest subject averaged regional bias 
($5.1\% \pm 2.6\%$, $p=0.003$ for {\petatlas} and $2.0\% \pm 1.5\%$, $p=0.006$ for {\petzte}).}
\nn{In addition, supplementary Figs. \ref{fig:tac_bias}, \ref{fig:R1_bias}, and \ref{fig:k2_bias}
show the bias in the time activity curves, and the $R_1$ and $k_2$ estimates in {\petatlas}
and {\petzte}, respectively.}

\section{Discussion}

Our analysis demonstrates that the bias caused by ZTEAC compared to CTAC as ground truth 
for brain PET/MR is small.
The magnitude of the maximum bias of 8\% is in agreement with the analysis of Sekine et 
al. \cite{Sekine2016b}.
In contrast to \cite{Sekine2016b}, we have evaluated more static PET
as well as dynamic PET acquisitions.
Moreover, the subjects in our analysis underwent a PET/MR protocol with realistic
PET acquisition times whereas \cite{Sekine2016b} only used an additional two minute PET/MR 
exam after a PET/CT acquisition. 

In contrast to the earlier evaluation of the quantitative accuracy of the AtlasAC 
\cite{Sekine2016,Sekine2016a} (range of VOI bias -5\% to +7.3\%) , 
our analysis showed that the AtlasAC implemented in
the GE SIGNA PET/MR can lead to \nn{individual} regional underestimations of up to 
-32\% (\nn{as observed in subject 24}).
A possible reason for the discrepancy is the fact that the number of subjects 
in \cite{Sekine2016,Sekine2016a} was much smaller compared to our study.
In this work, the biggest underestimations were found in a single subject \nn{(24)} where the
alignment of the atlas to the patient anatomy failed (see Fig.~\ref{fig:patex})
which caused a misclassification of some soft-tissue voxels as air voxels (pharynx) 
in the frontal region.
Since the atlas alignment is highly subject dependent, failures are hard to predict.
As demonstrated in subject 24, those failures can occur with the current implementation
of the AtlasAC leading to severe problems in regional quantification.

As observed in \cite{Sekine2016b}, another drawback of the current AtlasAC
is the fact that the introduced bias in the PET reconstruction shows a clear gradient
in the cranio-caudal direction.
Caudal VOIs such as the cerebellum (ca -4.2\%) and the anterior lateral temporal lobe (-4.2\%) 
show negative bias.
This is because part of the temporal and occipital bone are classified as soft tissue
in the current AtlasAC.
Moreover, there is a gross underestimation of the anterior part of the head including
the oropharynx, nasal cavities and cartilage tissue.

On the other hand, the superior cortical areas (frontal-parietal) show positive bias 
(+6.6\% in the left superior frontal gyrus).
This overestimation is caused by the fact that (a) the thickness of the superior skull seems
overestimated in the AtlasAC and (b) the atlas-based attenuation image is heavily post-smoothed such
that some soft tissue voxels in the superior gyri in the attenuation image are affected by spill 
over from skull voxels.

The cranio-caudal gradient in the bias distribution affects especially cerebral kinetic 
modeling analysis when using the cerebellum as the reference region.
This could be demonstrated in the kinetic modeling of the binding potential in the 
striatum of the \nnn{eleven} [$^{18}$F]PE2I subjects.
As a consequence of the observed negative bias in the cerebellum compared
to the striatum in {\petatlas} in the static cases,
AtlasAC leads to a small but systematic and significant 
overestimation of the binding potential of [$^{18}$F]PE2I in the striatum
(ca. $+5\%$ in the caudate nucleus and $+3.3\%$ in the putamen).
\nn{This positive bias can be understood by looking at Eq. (\ref{eq:srtm}).
Under the assumptions that $\mathrm{BP}_\mathrm{nd} \gg 1$ and  
$1 + \mathrm{BP}_\mathrm{nd} \gg R_1$ (which both are fullfilled for [$^{18}$F]PE2I in the striatum),
it can be seen that scaling $C_r(t)$ with $\alpha$ and at the same time scaling $R_1$, $k_2$,
and BP$_\mathrm{nd}$ with $\alpha^{-1}$ yields the exact same tissue response $C_t(t)$.
Since we can deduce from the analysis of the static examinations that $C_r(t)$ in the cerebellum 
is underestimated by ca. 4\% and that there is almost no bias in the striatum ($C_t(t)$), 
we would expect a 4\% overestimation in $R_1$, $k_2$, and BP$_\mathrm{nd}$ which is
in accordance with the results of the dynamic analysis as shown in Figs.~\ref{fig:BPnd_bias},
\ref{fig:R1_bias}, \ref{fig:k2_bias}.}

Using ZTEAC strongly reduces this bias in BP$_\mathrm{nd}$ (ca. $+2.0\%$ in the caudate 
nucleus and $+1.1\%$ in the putamen).
\nn{The performance of ZTEAC in the context of dynamic PET imaging is comparable
to the MaxProb multi atlas-based attenuation correction method.
In \cite{Merida2017},  Merida et al. could show that the MaxProb method leads to
a regional bias of -2\% to +5\% in the BP$_\mathrm{nd}$ of seven subjects examined with
[$^{18}$F]MPPF.}

It has been shown \cite{Sekine2016a,Burgos2015} and should be noted that the AtlasAC method
implemented in the GE SIGNA PET/MR is clearly outperformed by more advanced atlas-based methods.
Since the focus of this study was to analyze the performance of the ZTE-based attenuation
correction that will become clinically available on the SIGNA PET/MR,
a detailed analysis of more advanced atlas-based methods for MR-based attenuation correction
is beyond the scope of this study.

Compared to the detailed multi-center study of 11 methods for brain attenuation correction
for the Siemens mMR in 359 subjects by Ladefoged et al. \cite{Ladefoged2017},
it can be seen that the results for the regional quantitative accuracy of ZTEAC
are comparable to the best methods in \cite{Ladefoged2017} which showed a
global mean bias in the range of $-0.4\%$ to $+0.8\%$ with a standard deviation
of $1.2\%$ to $1.9\%$.
Also in terms of robustness (as seen in the standard deviation in the VOI-averaged bias)
and in terms of outlier behavior ZTEAC performs comparably to to best methods of \cite{Ladefoged2017}.
However, it should be noted that we could only analyze 30 subjects which influences the
detection of (rare) outliers.

Among the five best methods in \cite{Ladefoged2017} are three template-/atlas-based 
methods \cite{Izquierdo-Garcia2014,Burgos2015,Merida2015} and two ultra short echo time MR (UTE) 
segmentation-based methods \cite{Ladefoged2015,Juttukonda2015}.
Compared to the template-/atlas-based methods, the current ZTEAC for brain has the advantage 
that it does not rely on any anatomical prior information.
This might be beneficial in subjects with very abnormal brain anatomy (e.g. after surgery or 
traumatic brain injury) which needs further validation.

Finally, the fact that we had to exclude 6 out of \nn{48} patients \nn{(12.5\%)} due MR 
artifacts caused by dental implants demonstrates that there is a need for a reliable 
method for compensation of metal artifacts that can be applied in clinical routine. 

\nnn{A potential limitation of the study is the fact that the attenuation CTs used for CTAC
were acquired on a Siemens PET/CT system, but scaled to linear attenuation coefficients 
with the multi-linear scaling provided by GE.
This might lead to small residual uncertainties in the linear attenuation coefficients of the 
ground truth CTAC due to the fact that
the vendor-specific scaling procedures might be optimized for different effective x-ray 
spectra.
However, we do not expect this to be a major problem, because the multi-linear scaling 
curves of GE and Siemens are virtually identical up to 1200 HU.}

\section{Conclusion}
ZTE-based attenuation correction provides excellent quantitative accuracy for static 
and dynamic PET/MR imaging in all parts of the brain.
It is clearly superior to the Atlas-based head attenuation correction currently 
implemented in the GE SIGNA PET/MR and hereby obviates the major concern that was present in
the quantitative accuracy of brain PET/MR.

\section*{Conflict of interest}
GD is an employee of GE Healthcare, Cambridge, UK.
GS, MK, AR, NM, RP, JN, and KVL received travel grants from GE
Healthcare for attending workshops.
GS and KVL received travel grants for
presenting PET/MR results at GE user meetings.
KU Leuven received a research collaboration grant for partial funding of this
study.
GS is funded by the NIH project 1P41EB017183-01A1. 
AR and JN are funded by the Research Foundation Flanders (FWO) projects 12T7118N and G.0275.14N.
All other authors declare that they have no conflict of interest. \\

\bibliographystyle{jnm}
{\footnotesize\bibliography{manuscript}{}}       

\processdelayedfloats

\clearpage

\vspace*{\fill}
\huge{\bfseries\sffamily Supplementary materials}
\vspace*{\fill}

\renewcommand{\thefigure}{S\arabic{figure}}
\setcounter{figure}{0}
\renewcommand{\thetable}{S\arabic{table}} 
\setcounter{table}{0}
\renewcommand{\theequation}{S\arabic{equation}} 
\setcounter{equation}{0}

\begin{sidewaysfigure*}
  \centering
  \includegraphics[width = 1.0\textwidth]{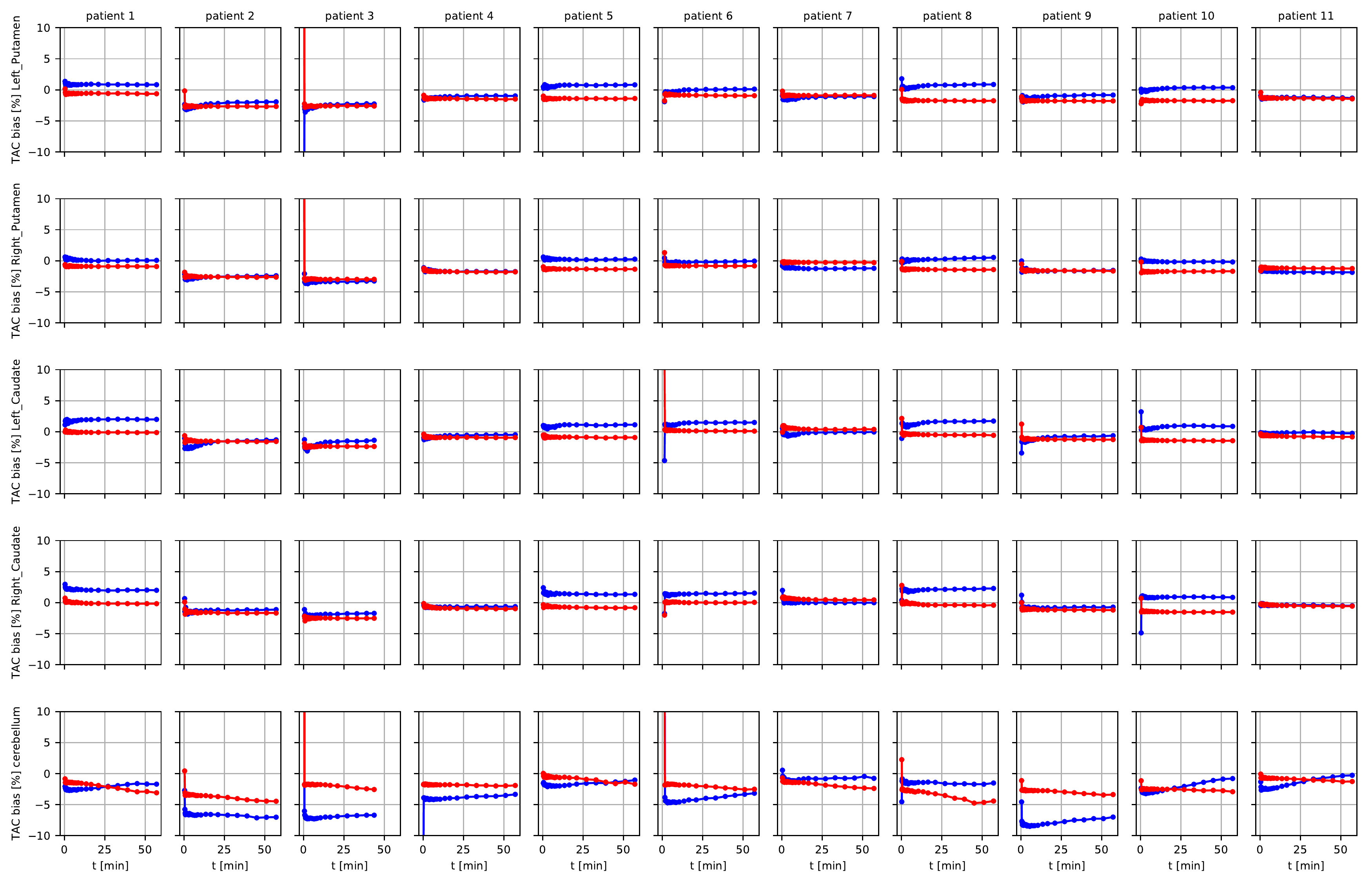}
  \caption{\nn{Bias in time activity curves of {\petatlas} (blue), and {\petzte} (red) 
               with respect to
               {\petct} in different regions for the 11 [$^{18}$F]PE2I subjects.}}
  \label{fig:tac_bias}
\end{sidewaysfigure*}

\begin{figure*}
  \centering
  \includegraphics[width = 0.95\textwidth]{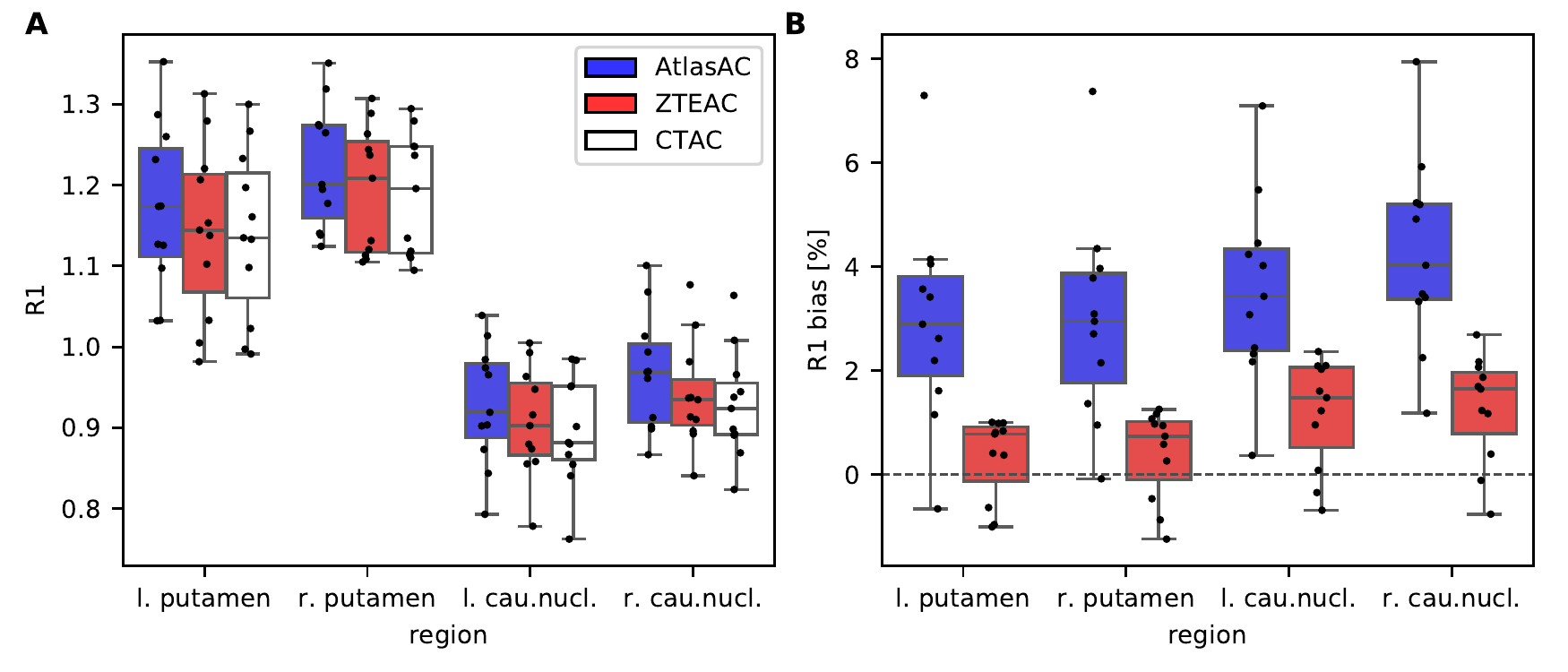}
  \caption{\nn{
           (A) Boxplot of $R_1$ values in four striatal regions of the 11 [$^{18}$F]PE2I subjects 
               obtained from {\petatlas}, {\petzte}, and {\petct}.   
           (B) Bias in $R_1$ estimation in {\petatlas}, {\petzte} compared to {\petct}.}}
  \label{fig:R1_bias}
\end{figure*}

\begin{figure*}
  \centering
  \includegraphics[width = 0.95\textwidth]{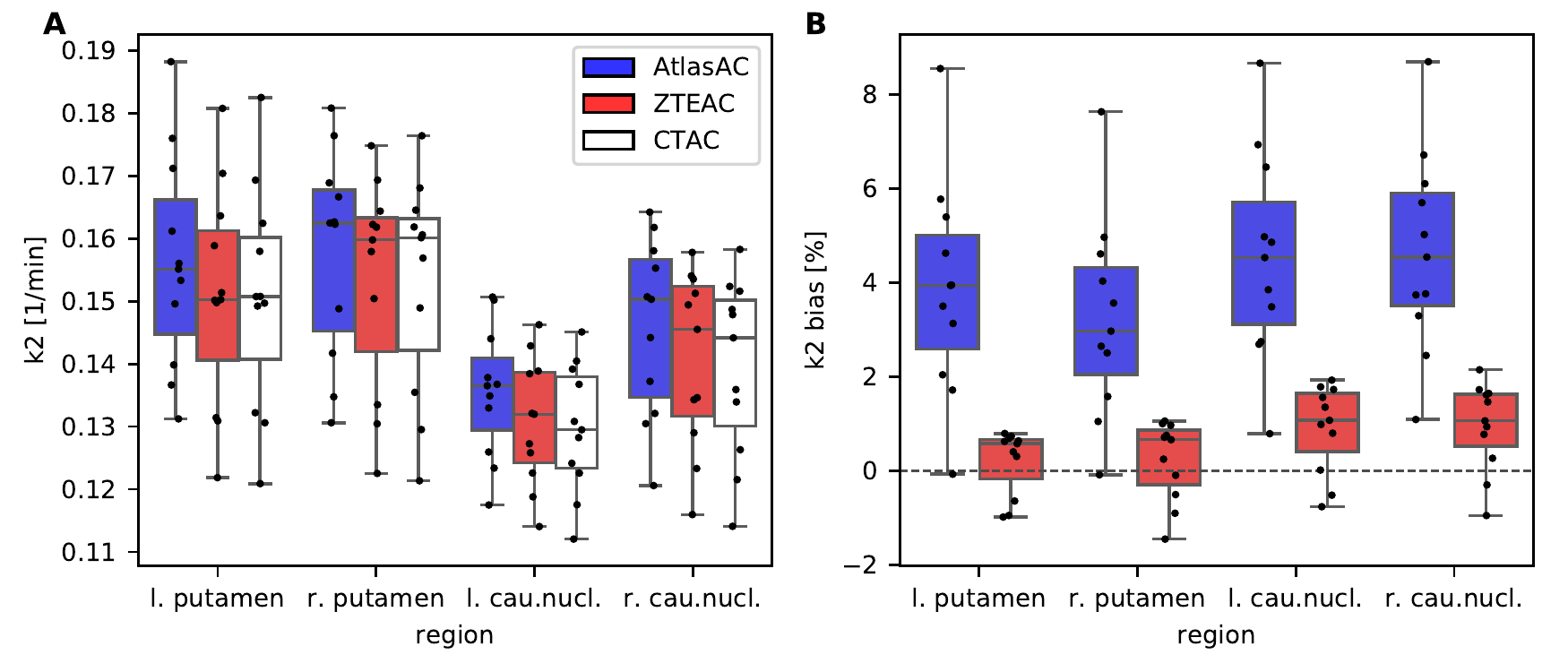}
  \caption{\nn{
           (A) Boxplot of $k_2$ values in four striatal regions of the 11 [$^{18}$F]PE2I subjects 
               obtained from {\petatlas}, {\petzte}, and {\petct}.   
           (B) Bias in $k_2$ estimation in {\petatlas}, {\petzte} compared to {\petct}.}}
  \label{fig:k2_bias}
\end{figure*}

\begin{sidewaystable*}
\footnotesize
\sffamily
\centering
\caption{Summary statistics of bias in the mean uptake in {\petatlas} and {\petzte}
         in all 85 anatomical VOIs averaged over all 30 static PET acquisitions.
         These data are also visualized in Fig.~\ref{fig:regional_bias}.
         \nn{Please note that the VOIs in the cerebellum had to be excluded in 
         3 subjects (6,7,21).}}
\label{tab:regional_bias}
\begin{tabular}{llrrrrrrrrrr}%
              &                  & \multicolumn{5}{c}{\bfseries bias in \petatlas [\%]} & \multicolumn{5}{c}{\bfseries bias in \petzte [\%]}  \\
\bfseries VOI & \bfseries region & \bfseries mean & \bfseries sd & \bfseries min & \bfseries max & \bfseries $p$ & \bfseries mean & \bfseries sd & \bfseries min & \bfseries max & \bfseries $p$ \\
\begin{filecontents*}{./rois1.csv}
ROI,region,mean,sd,min,max,zmean,zsd,zmin,zmax,apval,zpval
Amygdala left,medial temporal cortex,-3.1,3.6,-17.2,2.2,0.3,1.6,-3.3,3.2,<0.001,0.44
Amygdala right,medial temporal cortex,-2.5,2.9,-12.6,2.3,0.4,1.6,-2.8,3.5,<0.001,0.229
Anterior orbital gyrus left,frontal cortex,-1.5,4,-14.2,5.7,-0.3,2.5,-3.9,6.4,0.043,0.245
Anterior orbital gyrus right,frontal cortex,-2.2,3.7,-13.8,4,-0.5,2.5,-4.7,6,0.003,0.245
Anterior temporal lobe lateral part left,temporal cortex,-4.1,3.3,-11.8,1.3,-0.3,2.3,-4.2,4.6,<0.001,0.229
Anterior temporal lobe lateral part right,temporal cortex,-4.2,3.7,-12.3,2,-0.3,2.1,-5,3.9,<0.001,0.28
Anterior temporal lobe medial part left,temporal cortex,-3,4.2,-12.2,5.7,-0.5,2.6,-4.2,6,<0.001,0.088
Anterior temporal lobe medial part right,temporal cortex,-2.8,4.4,-13.9,5.4,-0.3,2.4,-4.3,4.9,0.001,0.105
Caudate nucleus left,striatum,0.9,2.3,-3.2,5.1,1.2,1.8,-2.5,4.2,0.119,0.005
Caudate nucleus right,striatum,1.7,2.6,-2.9,7.2,1.2,1.9,-2.2,4.2,0.009,0.015
Cerebellar white matter,-,-4.5,2.9,-10.9,1.5,1.1,2.7,-1.7,7.6,<0.001,0.211
Cerebellum left,cerebellum,-4.1,2.7,-10.6,0.8,0.4,2.3,-3.1,5.7,<0.001,0.859
Cerebellum right,cerebellum,-4.2,2.8,-10.6,2,0.7,2.4,-2.3,6.4,<0.001,0.455
Cerebral white matter left,cerebral WM,2.3,2.6,-4.5,7.6,-0.4,1.6,-3.5,2.6,<0.001,0.221
Cerebral white matter right,cerebral WM,2.2,2.6,-4.4,7.3,-0.3,1.6,-2.9,2.7,<0.001,0.28
Cingulate gyrus anterior part left,-,3.3,2.8,-2.2,9.4,-0.3,1.8,-3.8,2.7,<0.001,0.452
Cingulate gyrus anterior part right,-,3.7,2.9,-1.8,10.1,-0.3,1.8,-3.8,2.9,<0.001,0.477
Corpus callosum,-,2,2.4,-3.3,7,0.6,1.7,-2.4,3.7,<0.001,0.124
Cuneus left,occipital cortex,1,2.8,-4.9,5.7,-1.5,1.5,-4.5,1.2,0.077,<0.001
Cuneus right,occipital cortex,0.9,2.7,-5.1,5.1,-1.3,1.5,-4.6,1.3,0.124,<0.001
Fusiform gyrus left,temporal cortex,-3.9,4,-12.1,4.4,-0.4,2.6,-3.6,6.7,<0.001,0.092
Fusiform gyrus right,temporal cortex,-3.7,4,-11.1,6.1,0,2.5,-3.3,6.9,<0.001,0.318
Gyrus cinguli posterior part left,-,3.7,2.8,-3.4,9.9,0,1.6,-2.8,3.1,<0.001,0.903
Gyrus cinguli posterior part right,-,3.8,3,-3.2,10.9,0,1.7,-3.1,3.5,<0.001,0.746
Hippocampus left,medial temporal cortex,-2.4,2.6,-9.5,2.3,0.3,1.6,-2.5,3.5,<0.001,0.57
Hippocampus right,medial temporal cortex,-1.8,2.7,-10,2.8,0.2,1.6,-2.6,3.2,<0.001,0.73
Inferiolateral remainder of parietal lobe left,parietal cortex,2.7,3.2,-6.3,8.4,-0.9,1.5,-3.7,3.2,<0.001,0.003
Inferiolateral remainder of parietal lobe right,parietal cortex,1.6,3.2,-6.8,6.4,-0.5,1.7,-3.8,3.3,0.011,0.14
Inferior frontal gyrus left,frontal cortex,0.4,3.4,-9,7.7,-0.3,1.9,-3.1,3.5,0.641,0.349
Inferior frontal gyrus right,frontal cortex,-0.6,3.1,-7.2,6.4,0,2.1,-3.7,3.6,0.271,0.984
Insula left,-,-0.3,2.6,-8.4,4.3,0.1,1.5,-2.3,3.3,0.543,0.808
Insula right,-,-0.2,2.4,-6.8,3.8,0.2,1.7,-2.5,3.8,0.715,0.746
Lateral orbital gyrus left,frontal cortex,-1.6,3.7,-12.3,5.2,0,2.2,-2.5,5.5,0.016,0.685
Lateral orbital gyrus right,frontal cortex,-2.7,3.4,-11.3,4.6,0.1,2.5,-3.7,5.6,<0.001,0.919
Lateral remainder of occipital lobe left,occipital cortex,0.1,2.9,-4.5,7,-1.8,1.7,-4.9,1.5,0.919,<0.001
Lateral remainder of occipital lobe right,occipital cortex,0.1,2.8,-5.7,5.7,-1.6,1.6,-4.8,1.6,0.968,<0.001
Lateral ventricle (excluding temporal horn) left,-,2,2.7,-2.5,8.1,2.2,2.3,-2.8,5.7,0.004,<0.001
Lateral ventricle (excluding temporal horn) right,-,3.2,3,-2.2,11.2,1.7,2.3,-2.7,5.8,<0.001,0.001
Lateral ventricle temporal horn left,-,-2,2.6,-8.5,3.1,0.3,1.6,-2.5,3.4,<0.001,0.49
Lateral ventricle temporal horn right,-,-1.1,2.7,-9.1,4,0.2,1.7,-2.5,3.5,0.025,0.839
Lingual gyrus left,occipital cortex,-1.3,2.3,-6.3,3.9,-0.6,1.8,-3.3,3.7,0.002,0.016
Lingual gyrus right,occipital cortex,-1.2,2.2,-5.9,2.7,-0.5,1.6,-3.3,3.4,0.006,0.058
Medial orbital gyrus left,frontal cortex,-4,5.4,-27.7,3.6,0.1,2.1,-3.9,5.1,<0.001,0.952
\end{filecontents*}
\csvreader[head to column names]{./rois1.csv}{}%
{\\\ROI & \region & \mean & \sd & \min & \max & \apval & \zmean & \zsd & \zmin & \zmax & \zpval}%
\\
\end{tabular}
\end{sidewaystable*}

\begin{sidewaystable*}
\footnotesize
\sffamily
\centering
\caption{Continuation of Tab.~\ref{tab:regional_bias}}
\label{tab:regional_biastwo}
\begin{tabular}{llrrrrrrrrrr}%
              &                  & \multicolumn{5}{c}{\bfseries bias in \petatlas [\%]} & \multicolumn{5}{c}{\bfseries bias in \petzte [\%]}  \\
\bfseries VOI & \bfseries region & \bfseries mean & \bfseries sd & \bfseries min & \bfseries max & \bfseries $p$ & \bfseries mean & \bfseries sd & \bfseries min & \bfseries max & \bfseries $p$ \\
\begin{filecontents*}{./rois2.csv}
ROI,region,mean,sd,min,max,zmean,zsd,zmin,zmax,apval,zpval
Medial orbital gyrus right,frontal cortex,-3.8,4.8,-23.5,3.3,-0.2,2,-4,4.9,<0.001,0.516
Middle and inferior temporal gyrus left,temporal cortex,-3.1,3.1,-9.7,4,-0.1,2,-3.1,4.3,<0.001,0.339
Middle and inferior temporal gyrus right,temporal cortex,-3.7,3.1,-9.7,2.4,-0.2,1.9,-4,4.5,<0.001,0.245
Middle frontal gyrus left,frontal cortex,4.9,4.4,-4.8,16.1,-0.7,2.4,-8,3.1,<0.001,0.339
Middle frontal gyrus right,frontal cortex,3.6,4,-4.3,12.9,-0.5,2.3,-5.4,3.3,<0.001,0.49
Nucleus accumbens left,striatum,-1.8,2.8,-12.3,2.3,0.4,1.4,-2.6,3,0.001,0.129
Nucleus accumbens right,striatum,-1.1,2.4,-8,3.5,0.5,1.5,-3.1,3.2,0.021,0.067
Pallidum left,-,-1.1,2.2,-5.9,3,0.3,1.4,-2.3,2.7,0.01,0.349
Pallidum right,-,-0.7,2.1,-5.4,3.4,0.3,1.6,-2.6,3.6,0.1,0.393
Parahippocampal and ambient gyri left,-,-3.3,3.2,-10,3.3,0.1,1.9,-2.7,4.2,<0.001,0.67
Parahippocampal and ambient gyri right,-,-2.8,3.1,-10.5,3.9,0.1,2,-3.1,4.5,<0.001,0.685
Postcentral gyrus left,parietal cortex,5.2,3.9,-4.1,14.3,-0.1,1.8,-4.6,3.9,<0.001,0.7
Postcentral gyrus right,parietal cortex,4.5,3.6,-3.5,11.9,0,1.9,-3.5,3.8,<0.001,0.903
Posterior orbital gyrus left,frontal cortex,-2,3.3,-14.3,3.1,0.5,1.7,-2.3,5.3,<0.001,0.109
Posterior orbital gyrus right,frontal cortex,-2.1,2.8,-11.4,1.5,0.3,1.7,-2.7,4.2,<0.001,0.556
Posterior temporal lobe left,temporal cortex,-1.8,2.3,-5.9,3,-0.5,1.6,-3,2.6,<0.001,0.07
Posterior temporal lobe right,temporal cortex,-2.5,2.5,-7.5,2.1,-0.4,1.7,-3.6,3.4,<0.001,0.129
Precentral gyrus left,frontal cortex,6.2,4.1,-2.9,16.3,-0.2,2.1,-6.4,4,<0.001,0.7
Precentral gyrus right,frontal cortex,5.3,3.8,-2,13.5,-0.1,2.1,-5.1,4,<0.001,0.761
Pre-subgenual frontal cortex left,-,-1,2.2,-5.3,3.6,0.6,1.8,-2.5,4.2,0.015,0.141
Pre-subgenual frontal cortex right,-,-0.7,2,-4.1,3.1,0.7,1.7,-2.3,3.9,0.077,0.047
Putamen left,striatum,-0.8,2.4,-7.7,3.5,0.2,1.4,-2.4,2.5,0.05,0.503
Putamen right,striatum,-0.5,2.2,-6.1,3.7,0.3,1.6,-2.4,3.1,0.245,0.404
Straight gyrus left,frontal cortex,-4.2,5.7,-31.6,1.4,0.7,1.5,-2.9,3.2,<0.001,0.012
Straight gyrus right,frontal cortex,-3.7,4.6,-24.6,1.4,0.6,1.4,-2.8,3.3,<0.001,0.023
Subcallosal area left,-,-1.4,2.4,-7.8,2.6,0.9,1.5,-2.6,3.9,0.002,0.009
Subcallosal area right,-,-0.6,2.2,-5,2.9,0.9,1.5,-2.6,4.2,0.164,0.006
Subgenual frontal cortex left,-,-1.7,2.1,-6.4,2,0.9,1.5,-2.5,3.6,<0.001,0.006
Subgenual frontal cortex right,-,-1.2,1.9,-4.5,2.5,1,1.5,-2.1,3.3,0.002,0.004
Substantia nigra left,-,-1.9,1.9,-5.7,1.3,0.1,1.6,-3.4,3.2,<0.001,0.761
Substantia nigra right,-,-1.5,2.2,-7.8,2.5,0.3,1.5,-2.5,3.5,<0.001,0.67
Superior frontal gyrus left,frontal cortex,6.6,4,-1.7,16.5,-0.5,2.4,-7.6,3.6,<0.001,0.416
Superior frontal gyrus right,frontal cortex,6.2,3.8,-1.2,15.2,-0.5,2.4,-7.3,3.8,<0.001,0.44
Superior parietal gyrus left,parietal cortex,5.2,3.4,-3.6,11.9,-0.7,1.5,-3,3.2,<0.001,0.008
Superior parietal gyrus right,parietal cortex,5,3.3,-2.9,11.7,-0.5,1.6,-3.2,3.4,<0.001,0.035
 Superior temporal gyrus anterior part left,temporal cortex,-3,3.5,-13.2,2.7,0.2,2.2,-4.2,6.2,<0.001,0.855
 Superior temporal gyrus anterior part right,temporal cortex,-3.1,3.3,-12.4,3.3,0.3,2.4,-4.3,7,<0.001,0.984
Superior temporal gyrus posterior part left,temporal cortex,-0.9,2.9,-9.1,4.9,-0.1,1.5,-2.9,2.9,0.052,0.598
Superior temporal gyrus posterior part right,temporal cortex,-1.8,2.6,-8.9,3.2,0.1,1.8,-3.1,3.5,<0.001,0.919
Thalamus left,thalamus,-0.2,2.1,-4.6,3.3,0.4,1.6,-2.3,3.4,0.685,0.271
Thalamus right,thalamus,0.1,2.2,-5.1,3.4,0.4,1.7,-2.4,3.5,0.746,0.44
Third ventricle,-,-0.4,2.3,-5.3,4,1,1.8,-2,5.4,0.152,0.009
\end{filecontents*}
\csvreader[head to column names]{./rois2.csv}{}%
{\\\ROI & \region & \mean & \sd & \min & \max & \apval & \zmean & \zsd & \zmin & \zmax & \zpval}%
\\
\end{tabular}
\end{sidewaystable*}

\begin{table*}
\footnotesize
\sffamily
\centering
\caption{Summary statistics of bias in the mean uptake in {\petatlas} and {\petzte}
         in all 30 static PET acquisitions averaged over all VOIs.
         These data are also visualized in Fig.~\ref{fig:subjects_box}.
         \nn{Please note that the VOIs in the cerebellum had to be excluded in 
         3 subjects (6,7,21).}}
\label{tab:subjects}
\begin{tabular}{lrrrrrrrrrr}%
              & \multicolumn{5}{c}{\bfseries bias in \petatlas [\%]} & \multicolumn{5}{c}{\bfseries bias in \petzte [\%]}  \\
\bfseries subject & \bfseries mean & \bfseries sd & \bfseries min & \bfseries max & \bfseries $p$& \bfseries mean & \bfseries sd & \bfseries min & \bfseries max & \bfseries $p$\\
\begin{filecontents*}{./subjects.csv}
subject,mean,sd,min,max,zmean,zsd,zmin,zmax,apval,zpval
1,-0.7,3,-7,4.9,1.7,1,-0.8,3.5,0.056,<0.001
2,0,2.5,-7.3,6.6,2.9,1.8,-1,7.6,0.835,<0.001
3,-0.8,3.1,-6,6.5,-0.3,1.1,-2.7,2.5,0.014,0.02
4,0.7,2.8,-4.1,7.1,-1.7,1.2,-4.3,1.9,0.273,<0.001
5,-0.4,2.7,-4.8,7.6,-0.9,1.2,-3.6,2.3,0.02,<0.001
6,0.6,2.7,-3.7,8,0,1.5,-3.4,3.2,0.601,0.952
7,1.6,2.3,-3,7.2,-0.7,1.4,-3.7,3,<0.001,<0.001
8,-0.7,2.7,-6.6,5.7,-1.1,1.1,-4.7,1,0.045,<0.001
9,0.1,3.2,-4.6,7.5,-1.6,1.8,-8,0.4,0.602,<0.001
10,-2.9,1.8,-6.5,2.5,-2.4,0.7,-4.2,-0.9,<0.001,<0.001
11,-4,2.8,-9.9,2.1,-1.5,1,-3.6,0.8,<0.001,<0.001
12,0.8,3.3,-5.2,9.5,0.6,1.1,-2.3,2.5,0.27,<0.001
13,-0.2,2.5,-4.7,5.2,-0.4,1.1,-3.5,3.3,0.318,<0.001
14,-1.6,3.1,-6.3,6.5,0.6,0.7,-1.3,3.4,<0.001,<0.001
15,1,3,-3.7,8.1,2.1,1.3,-0.7,5.9,0.095,<0.001
16,-1.6,2.7,-7.9,3.8,0,0.9,-2,2.9,<0.001,0.733
17,0.2,2.6,-4.4,6.8,0.6,0.9,-2,3.6,0.813,<0.001
18,2.7,3.1,-1.8,10.2,2.6,1.1,0.6,6.8,<0.001,<0.001
19,4,4.2,-3.3,13.9,0.5,1.3,-1.8,5.7,<0.001,<0.001
20,2.4,2.6,-3.3,8.7,1.1,1,-1.7,4.5,<0.001,<0.001
21,3,4.3,-2.4,16.3,1,1.8,-2.8,6.9,<0.001,<0.001
22,4.5,3.9,-1,16.5,2.7,1.6,-1.9,7,<0.001,<0.001
23,-0.6,3.7,-6.8,7.7,-0.9,1.1,-3.2,3.3,0.054,<0.001
24,-7.8,5.6,-31.6,-1.2,0.1,1.2,-2.6,2.5,<0.001,0.878
25,-2.5,4,-9.3,7.4,-1.6,1.1,-4.7,0.9,<0.001,<0.001
26,-3.2,3.6,-9.6,7.6,-2.2,1.4,-5,1.5,<0.001,<0.001
27,1.6,3.5,-3.3,9.7,1.2,0.9,-0.4,4.4,0.001,<0.001
28,-3.8,4.6,-12.2,4.6,-0.9,1.2,-3.7,2,<0.001,<0.001
29,-4.8,3,-11.6,1.5,-2.5,0.7,-4.9,-1.2,<0.001,<0.001
30,1.3,4.6,-7.5,9.9,2.8,1,0.6,5.8,0.02,<0.001
\end{filecontents*}
\csvreader[head to column names]{./subjects.csv}{}%
{\\\subject & \mean & \sd & \min & \max & \apval & \zmean & \zsd & \zmin & \zmax & \zpval}%
\\
\end{tabular}
\end{table*}

\begin{table*}
\footnotesize
\sffamily
\centering
\caption{Summary statistics of bias in the mean uptake in {\petatlas} and {\petzte}
         in different anatomical regions averaged over all 30 static PET acquisitions.
         These data are also visualized in Fig.~\ref{fig:regions_box}.
         \nn{Please note that the VOIs in the cerebellum had to be excluded in 
         3 subjects (6,7,21).}}
\label{tab:region}
\begin{tabular}{lrrrrrrrrrr}%
              & \multicolumn{5}{c}{\bfseries bias in \petatlas [\%]} & \multicolumn{5}{c}{\bfseries bias in \petzte [\%]}  \\
\bfseries region & \bfseries mean & \bfseries sd & \bfseries min & \bfseries max & \bfseries $p$ & \bfseries mean & \bfseries sd & \bfseries min & \bfseries max & \bfseries $p$  \\
\begin{filecontents*}{./regions.csv}
region,mean,sd,min,max,zmean,zsd,zmin,zmax,zpval,apval
frontal cortex,0.3,5.6,-31.6,16.5,-0.1,2.1,-8,6.4,0.408,0.777
temporal cortex,-3,3.5,-13.9,6.1,-0.2,2.1,-5,7,<0.001,<0.001
parietal cortex,4,3.7,-6.8,14.3,-0.4,1.7,-4.6,3.9,<0.001,<0.001
occipital cortex,0,2.7,-6.3,7,-1.2,1.7,-4.9,3.7,<0.001,0.461
medial temporal cortex,-2.4,3,-17.2,2.8,0.3,1.6,-3.3,3.5,0.151,<0.001
striatum,-0.3,2.7,-12.3,7.2,0.6,1.6,-3.1,4.2,<0.001,0.032
thalamus,-0.1,2.2,-5.1,3.4,0.4,1.6,-2.4,3.5,0.177,0.956
cerebellum,-4.2,2.7,-10.6,2,0.6,2.4,-3.1,6.4,0.541,<0.001
cerebral WM,2.3,2.6,-4.5,7.6,-0.3,1.6,-3.5,2.7,0.103,<0.001
\end{filecontents*}
\csvreader[head to column names]{./regions.csv}{}%
{\\\region & \mean & \sd & \min & \max & \apval & \zmean & \zsd & \zmin & \zmax & \zpval}%
\\
\end{tabular}
\end{table*}

\begin{table*}
\footnotesize
\sffamily
\centering
\caption{BP$_\mathsf{nd}$ derived from {\petct} and bias of BP$_\mathsf{nd}$ in {\petatlas} and {\petzte} for the eight
         [$^{18}$F]PE2I acquisitions in the striatum using {\petct} as the ground truth.}
\label{tab:bpbias}
\begin{tabular}{lrrrrrrrrrrrr}%
              & \multicolumn{2}{c}{\bfseries $\text{BP}_\text{nd,CTAC}$} & \multicolumn{5}{c}{\bfseries bias in $\text{BP}_\text{nd,AtlasAC}$ [\%]} & \multicolumn{5}{c}{\bfseries bias in $\text{BP}_\text{nd,ZTEAC}$ [\%]}  \\
\bfseries region & \bfseries mean & \bfseries sd & \bfseries mean & \bfseries sd & \bfseries min & \bfseries max & \bfseries $p$ & \bfseries mean & \bfseries sd & \bfseries min & \bfseries max & \bfseries $p$ \\
\begin{filecontents*}{./pe2i_bpnd.csv}
region,bpmean,bpstd,bamean,bastd,bamin,bamax,pa,bzmean,bzstd,bzmin,bzmax,pz,pc
left caudate nucleus,3.3,0.5,5.0,2.6,1.0,9.4,0.003,2.0,1.5,-0.1,4.6,0.008,0.008
left putamen,4.3,0.5,3.7,2.7,-0.4,8.7,0.008,1.0,1.1,-0.6,2.7,0.026,0.013
right caudate nucleus,3.3,0.6,5.1,2.6,1.1,9.3,0.003,2.0,1.5,-0.3,4.8,0.006,0.008
right putamen,4.3,0.4,3.0,2.5,-0.6,7.8,0.008,1.1,1.3,-0.8,3.1,0.033,0.026
all combined,3.8,0.7,4.2,2.6,-0.6,9.4,<0.001,1.5,1.4,-0.8,4.8,<0.001,<0.001
\end{filecontents*}
\csvreader[head to column names]{./pe2i_bpnd.csv}{}%
{\\\region & \bpmean & \bpstd & \bamean & \bastd & \bamin & \bamax & \pa & \bzmean & \bzstd & \bzmin & \bzmax & \pz}%
\\
\end{tabular}
\end{table*}


\end{document}